\documentclass[aps,prl,superscriptaddress,twocolumn,showpacs,amsfonts]{revtex4}
\usepackage{graphicx,amsfonts}
\usepackage{float}

\begin{document}

\title{MD simulations and continuum theory of partially fluidized shear granular flows}

\author{Dmitri Volfson}
\affiliation{Institute for Nonlinear Science, University of California,
San Diego, La Jolla, California 92093-0402 }
\author{Lev S. Tsimring}
\affiliation{Institute for Nonlinear Science, University of California,
San Diego, La Jolla, California 92093-0402 }
\author{Igor S. Aranson}
\affiliation{Argonne National Laboratory,
9700 South  Cass Avenue, Argonne, Illinois 60439}

\date{\today}

\begin{abstract}
We carry out a detailed comparison of soft particle molecular
dynamics simulations with the theory of partially fluidized
shear granular flows.
We verify by direct simulations a constitutive relation based on the separation of the
shear stress tensor into a fluid part proportional to the strain rate tensor,
and a remaining solid part.
The ratio of these two components is determined by the order parameter.
Based on results of the simulations we construct the ``free energy''
function for the order parameter.
We also present the simulations
of the stationary deep 2D granular flows
driven by an upper wall and compare it with the continuum theory.
\end{abstract}
\pacs{45.70.Cc, 46.25.-y,83.80.Fg}
\date{\today}

\maketitle

When the ratio of shear to normal stress in a packed granular matter
exceeds a certain threshold value, the granular matter yields and a flow
ensues.  In the last few years there have been many experimental,
numerical and theoretical studies
\cite{behringer99,gollub98,gollub00,lemaitre,starton,pouliquen99,daerr,Tsai02,grest91,wolf96,grest02,sparks99}
that explored a broad range of granular flow conditions.  While dilute
granular flows can be well described by the kinetic theory of
dissipative granular gases \cite{Jenkins85}, dense granular flows still
present significant difficulties for theoretical description.  A
continuum theory of slow dense granular flows based on the so-called
associated flow rule that relates the strain rate and the shear stress
was proposed in Ref.  \cite{savage98}.  This description neglects the
effects of the dry friction between the grains and works only in a
fluidized state, so it cannot not describe hysteretic nature of the
granular flow relevant for stick-slips \cite{gollub98}, avalanching
\cite{daerr}, etc.  A similar model based on a Newtonian stress-strain
constitutive relation with density dependent viscosity was proposed in
Refs.  \cite{gollub00}.  In this model, the viscosity  diverges at the
fluidization threshold when the density approaches the random close
packing density of grains.

Recently  we proposed a phenomenological order parameter (OP)
description of the fluidization transition \cite{AT1}. The OP specifies
the ratio between solid and fluid parts of the stress tensor.  The
viscosity is defined as a ratio of the {\em fluid} part of the shear
stress to the strain rate and remains finite at the fluidization
threshold.  This model yielded a good qualitative description of  broad
variety of phenomena occurring in granular flows.

In this Letter  we report  on  2D soft particle molecular dynamics
simulations performed to validate and quantify our OP theory.  To
fit the equation for the OP and stress-strain relation we
performed simulations of the granular flow in a thin Couette
geometry. The obtained set of equations was used to calculate the
stress and velocity profiles in a {\em different} system, a thick
granular layer under non-zero gravity driven by a moving heavy
upper plate.

The theory \cite{AT1} is based on a standard momentum conservation
equation and the incompressibility condition applicable for slow dense flows.
To close the system, we assumed that
the stress tensor $\sigma$ is comprised of two parts,
a solid part $\sigma^s$, and a fluid part $\sigma^f$ (taken in a
purely Newtonian form)
\begin{equation}
\sigma^f_{ij}=p_f\delta_{ij}-\mu_f\dot\gamma_{ij}
\label{sigmaf}
\end{equation}
where $p_f$ is the isotropic ``partial'' fluid pressure, $\mu_f$ is the viscosity coefficient
associated with the fluid stress tensor.
We set the fluid part of
off-diagonal  components of the stress tensor to be proportional to
the off-diagonal components of the full stress tensor with the
proportionality  coefficient being a function of the OP $\rho$,
\begin{equation}
\sigma^f_{yx}=q(\rho)\sigma_{yx}; \; \sigma^s_{yx}=(1-q(\rho))\sigma_{yx}.
\label{sigmaf1}
\end{equation}
We choose a fixed range for the OP such that it is zero in a
completely fluidized state and one in a completely static regime.
Thus, the function $q(\rho)$ has the property $q(0)=1,\ q(1)=0$.
In Refs.\cite{AT1} for simplicity we postulated $q(\rho)=1-\rho$.
A similar relationship can be introduced for the diagonal terms of
the stress tensor
\begin{eqnarray}
\sigma^f_{xx} &=& q_x(\rho)\sigma_{xx},\ \sigma^f_{yy} =
q_y(\rho)\sigma_{yy}
\label{sigmaf2}\\
\sigma^s_{xx} &=& (1-q_x(\rho))\sigma_{xx},\ \sigma^s_{yy} =
(1-q_y(\rho))\sigma_{yy}
\label{sigmas2}
\end{eqnarray}

The dynamics of the OP was assumed to be relaxational in nature and
controlled by the generic Ginzburg-Landau equation,
\begin{equation}
\frac{D\rho}{Dt}=D\nabla^2\rho-F(\rho,\delta)
\label{GL}
\end{equation}
Here $D/Dt$ is the material derivative, $D$ is the diffusion coefficient,  
$F(\rho,\delta)$ is the derivative of
the free energy density which has a quartic polynomial
form to account for the bistability near the solid-fluid transition.
Control parameter $\delta$
was taken to be a
linear function of $\phi=\max |\sigma_{mn}/\sigma_{nn}|$.

The OP  $\rho$ which plays a pivotal role in the theory, should be
associated with the "microscopic" properties of the granular
assembly. At any moment of time all contacts among the grains can
be classified as either {\em fluid-like} or {\em solid-like}. A
contact is considered fluid-like if two particles slide past each
other or briefly collide, and is solid-like if two particles are
jammed together for longer than a characteristic collision time.
Here we postulate that the OP  can be introduced as a ratio
between space-time averaged numbers of solid contacts
$\overline{\langle Z_s\rangle}$ and all contacts
$\overline{\langle Z\rangle}$ within a sampling area,
\begin{equation}
\rho(y)=\overline{\langle Z_{s}\rangle} / \overline{\langle Z\rangle} .
\label{rho}
\end{equation}
where $\langle \xi \rangle$ and $\overline{\xi}$ stand for
averaging of $\xi$ in space and time respectively. This definition
satisfies both limiting cases: when a granulate is in a solid
state all contacts are stuck and $\rho = 1$; when it is strongly
agitated $\overline{\langle Z_s\rangle}$ is zero and
$\overline{\langle Z\rangle}$ is small but finite, therefore $\rho
= 0$. Our  OP is expected to be sensitive to the degree of
fluidization. A small rearrangement of the force network may
result in strong fluctuations of $\rho$, while  the solid fraction
and  granular temperature remain virtually constant. This quantity
is difficult to measure in experiments, however it can be found in
soft-particle molecular dynamics.

{\it Molecular Dynamics Simulations} \cite{note1}. The grains are
assumed to be non-cohesive, dry, inelastic disk-like particles.  Two
grains interact via normal and shear forces whenever they overlap.  For
the normal impact we employed \emph{spring-dashpot} model \cite{wolf96}.
This model accounts for repulsion and dissipation; the repulsive
component is proportional to the degree of the overlap and the velocity
dependent damping component simulates the dissipation.  The model for
shear force is based upon technique developed in \cite{cundall79}. The
motion of a grain is obtained by integrating the Newton's equations with
forces and torques produced by interactions with neighbors  and walls.
For detailed discussion of the advantages and limitations of the
employed model see Refs. \cite{wolf96,grest91,grest02}.  The
computational domain spans $L_x\times L_y$ area, and is periodic in
horizontal direction $x$.  The grain diameters are uniformly distributed
around mean with relative width $\Delta_r$ to avoid crystallization
effects \cite{grest91}.  The material parameters of grains were chosen
similar to Ref.\cite{grest02}.  All  quantities are normalized by an
appropriate combination of the average particle diameter $d$, mass $m$,
and gravity $g$.

\begin{figure}[ptb]
\includegraphics[width=3.5in]{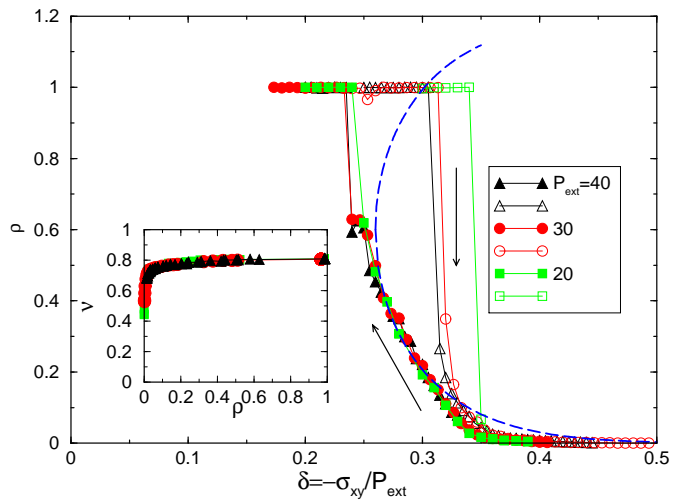}
\caption{$\rho$ vs $\delta$  for $P_{ext}=20$ ($\Box$),30
($\bigcirc$) and 40 ($\bigtriangleup$). Open symbols correspond to
ramp-up, and filled symbols to ramp-down in shear stress. Dashed
line shows fit Eq. \protect \ref{FF}. Inset: density $\nu$ vs
$\rho$ for the same three $P_{ext}$.} \label{norm}
\end{figure}

We studied a thin ($50\times 10$) granular layer sandwiched
between two ``rough plates'' under fixed external pressure $P_{ext}$ and zero gravity conditions.
The rough plates were simulated by two straight
chains of large grains ``glued'' together.  Two opposite forces
${\bf F}_1=-{\bf F}_2$ were applied to the plates along the horizontal
$x$ axis to induce shear stress in the bulk.  We started with weak forces
and slowly ramped them up in small increments.
After that we ramped
the shear forces down until the granular layer was jammed again. At
every ``stop'' we measured all stress components, strain rate, and the
OP by averaging over the whole layer and over time. These simulations were
carried out at several values of the external pressure $P_{ext}$.
Figure \ref{norm} shows $\rho$ as a function
of the normalized  shear stress $\delta=-\sigma_{yx}/P_{ext}$.
With this normalization, the results of several simulations with different
$P_{ext}$ collapse on a single bifurcation curve.
We observe a quiescent state $\rho=1$  until $\delta$  reaches a certain critical
value $\delta_1\approx 0.32$. This value differs slightly for different
runs because of the finite system size and absence of self-averaging in
the static regime. Above $\delta_1$, $\rho$ abruptly drops to
approximately $0.15$.  At larger $\delta$, the OP rapidly approaches zero.
The return curve corresponding to the diminishing of the shear stress
follows roughly the same path, and then continues to another (smaller)
value of the shear stress, $\delta_2\approx 0.23$. At this point
the OP jumps back to one, and the granular layer returns to a jammed
state.
The striking feature of this bifurcation diagram is the hysteretic behavior
of the OP as a function
of the shear stress. This hysteresis was anticipated in our model
\cite{AT1}, and now we are in a position to describe it
quantitatively.  Assuming that there is an (unobserved) unstable
branch of the bifurcation curve which merges with the stable branch
at $\delta\approx \delta_1$, we make a simple analytic fit,
\begin{equation}
F(\rho,\delta)=(1-\rho)
\left(\rho^2-2\rho_*\rho+\rho_*^2\exp[-A(\delta^2-\delta_*^2)]\right)
\label{FF}
\end{equation}
with $\rho_*=0.6, A=25,\delta_*=0.26$ (see Figure \ref{norm},
dashed line) and use it in equation (\ref{GL}).  The inset of
Figure 1 depicts the density $\nu$ vs. the OP $\rho$ for the same
runs. The  density  stays almost constant in a wide range of the
OP $0.1<\rho<1$. This shows that unlike the particle density, our
OP  is a sensitive characteristic of slow dense granular flows
reflecting subtle changes in the contact network and  the
structure of the stress distribution.

We also probed the relaxation dynamics of the OP by studying the
response of the system on small perturbation in the the hysteretic
region, $\delta_2<\delta<\delta_1$ \cite{note1}.  From these
simulations we find that the intrinsic time scale of the  OP
relaxation is rather small, $O(1)$.  Our thin Couette flow system
did not allow us to probe the local coupling of the OP  since
$\rho\approx const $ throughout the system.  In the absence of
such data this coupling was modeled by the linear diffusion term
in (\ref{GL}) with $D=const$.

\begin{figure}[ptb]
\includegraphics[angle=270,width=3.5 in]{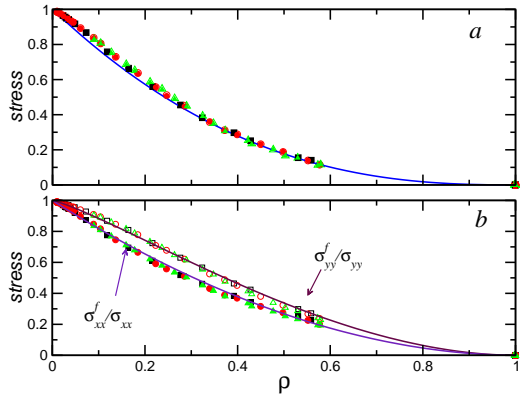}
\caption{
Ratios of the fluid stress components to the  full
stress components
$\sigma^{f}_{ij}/\sigma_{ij}$ vs. $\rho$ for $P_{ext}=20$
($\Box$),30 ($\bigcirc$) and
40 ($\bigtriangleup$). a -
shear stress components $\sigma_{yx}$ (open symbols),  $\sigma_{yx}$
(filled symbols), line is a fit
$q(\rho)=(1-\rho)^{2.5}$, b - normal stress components
$\sigma_{xx}$ (closed symbols),
$\sigma_{yy}$ (open symbols), lines are the fits $q_x(\rho)=(1-\rho)^{1.9},\
q_y(\rho)=(1-\rho^{1.2})^{1.9}$.
}
\label{stress30}
\end{figure}

{\it The constitutive relation} was fitted
using the same Couette flow simulations. 
We analyzed the fluid stress $\sigma^f_{ij}$ and the solid
stress $\sigma^s_{ij}$ separately during our ramp-down simulations at
three different values of $P$. Figure \ref{stress30} shows the ratios of
fluid tensor components to the corresponding full tensor components as
functions of $\rho$.  We observe that data for
$\sigma^s_{xy}/\sigma_{xy}$ and $\sigma^s_{yx}/\sigma_{yx}$ for different
$P$ fall onto a single curve which is fitted by $q(\rho)=(1-\rho)^{2.5}$
(Figure \ref{stress30},a).
Repeating the  procedure with diagonal
elements  yields different scaling, see
Figure \ref{stress30}, b. A small but noticeable difference is
evident between $\sigma^f_{xx}/\sigma_{xx}$ and $\sigma^f_{yy}/\sigma_{yy}$.
Detailed analysis shows that in fact fluid parts of the diagonal
components of the stress tensor $\sigma^f_{xx}$ and $\sigma^f_{yy}$ are
nearly identical \cite{note2}, and the difference is mainly due to the
solid part of the normal stresses.
Functions $q_{x,y}(\rho)$  approach 1 as $\rho\to 0$, but
they may have different functional form to reflect the
anisotropy of the solid stress tensor.
In our Couette flow, $\sigma_{xx}$ and $\sigma_{yy}$ can
be  fitted by $q_x(\rho)\approx (1-\rho)^{1.9}$ and
$q_y(\rho)\approx (1-\rho^{1.2})^{1.9}$, respectively, see
Figure \ref{stress30},b.  We observe that even
in a partially fluidized regime, the ``fluid phase'' component
behaves as a fluid with a well-defined isotropic ``partial'' pressure $p_f$
which is zero in a solid state  and is becoming the full
pressure in a completely fluidized state.

\begin{figure}[ptb]
\includegraphics[angle=270,width=3.5in]{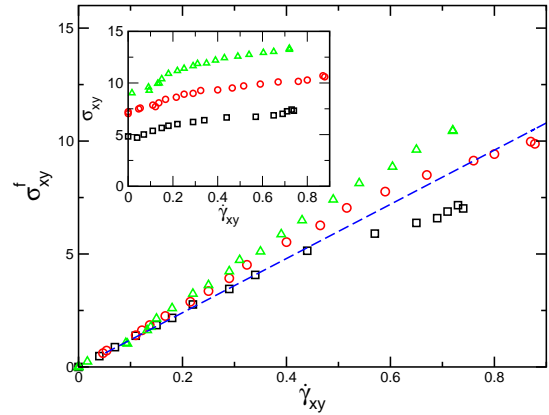}
\caption{Stress-strain rate relation for a thin granular Couette flow at
$P_{ext}=20$ ($\Box$),30 ($\bigcirc$) and
40 ($\bigtriangleup$).
Fluid shear stress vs strain
rate, the straight line is a constant viscosity fit
$\sigma^f=12\dot\gamma$;
Inset: full shear stress vs strain rate }
\label{visc}
\end{figure}

To test the stress-strain relation
(\ref{sigmaf}) we plot $-\sigma^f_{yx}$ vs $\dot\gamma_{yx}$,
see Figure \ref{visc}. At small $\dot\gamma_{yx}$ all
curves are close to the same straight line $\sigma^f_{yx}\approx
12\dot\gamma_{yx}$, i.e.  the Newtonian scaling for fluid shear stress
holds reasonably well. The deviations at large $\dot\gamma_{yx}$ are
evidently caused by variations of density and temperature in the dilute
regime as to be expected from kinetic theory of dilute granular flows
\cite{Jenkins85}.  The full shear stress, of course, does not vanish as
$\dot\gamma_{yx}\to 0$ (Figure \ref{visc}, inset). Therefore, the
standard viscosity coefficient defined as the ratio of the full shear
stress and strain rate diverges at the fluidization threshold as
observed in Ref. \cite{gollub00}.

We applied our  theoretical description which was
formulated above on the basis of MD simulations of a
thin Couette flow with no gravity, to  a
{\it different} system, a shear granular flow in a thick granular
layer under gravity
driven by the upper plate which was pulled with constant speed
$V$ (or constant force $F$).  A similar system has been studied
experimentally Refs.  \cite{gollub98,Tsai02}.
We simulated up to 20,000 particles in a periodic rectangular box under a heavy
plate which was moved horizontally  in $x$-direction.
\begin{figure}[ptb]
\includegraphics[angle=0,width=3.in]{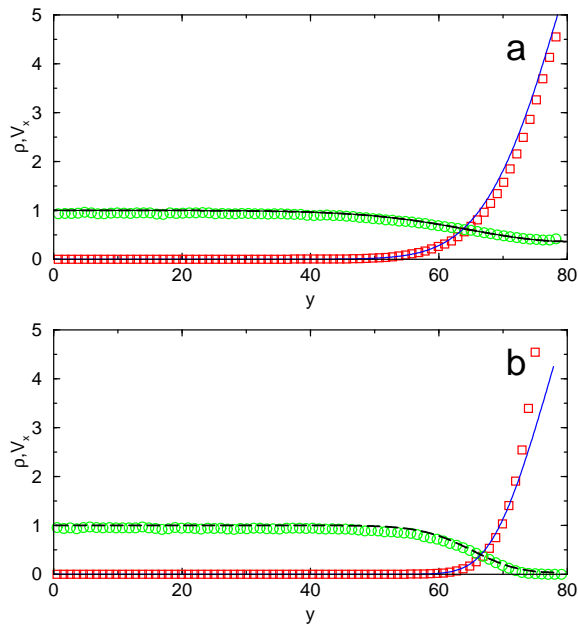}
\caption{Vertical profiles of $\rho$  and  $V_x$ in a thick
granular layer driven at the surface by a heavy plate for
5,000 particles, in the box  $L_x=50,L_y=100$, a -
$P_{ext}=50$, pulling velocity $V=5$, $D=5$, b - $P_{ext}=10$, pulling
velocity $V=50$, $D=1$.
Lines show the theoretical results obtained from the continuum
model \protect(\ref{GL}), (\ref{FF}), open symbols show numerical
data. }
\label{profiles}
\end{figure}
We systematically carried out comparison between MD simulations and the
continuum theory using  the stress-strain relations and the specific
form of the OP equation described above. We used no-flux boundary conditions
for the OP and no-slip condition for the velocity at the bottom plate.
The boundary condition at the top plate where the flow can be in a
dilute regime, is a separate issue which we do not address here. We
assumed that the shear stress component in the bulk is specified by
applied force, and calculated the velocity profile using the constitutive
relation (\ref{sigmaf})-(\ref{sigmas2}).  We found that the constitutive
relations determined from the thin Couette experiment hold for this
system as well. Selected results for the OP and velocity profiles are
presented in Figure \ref{profiles}.  As seen from the Figure, the
vertical profiles of the OP  and the horizontal velocities are
reasonably well described by the theory. However, for low pressure the
horizontal velocity profiles deviate from the numerical data for low
pressure runs, apparently because the viscosity coefficient is no longer
a constant in a dilute region near the top plate.  The only fitting
parameter used was the diffusion constant $D$ in the OP equation, which
has not been determined in our simulations of the thin layer.  It
appears that the diffusion coefficient depends on the applied pressure
and the strain rate, however more detailed numerical experiments are
needed. 

In conclusion,  we calibrated the theory of partially granular
flows \cite{AT1} on the basis of a series of 2D soft particle molecular dynamics simulations.
The OP which controls the fluidization transition, was
defined as the fraction of solid-like contacts among
particles. Measurements of the OP, the stress tensor, and the
strain rate in a thin Couette cell allowed us to quantify the
constitutive relations based on the relaxational OP dynamics.
We studied the flow structure of a thick
surface driven granular flow under gravity and  found the model predictions
to be in a good {\em quantitative} agreement with soft-particle MD simulations.
Our results support an intriguing
interpretation for the OP  description of dense and slow granular flows.
The granular matter under shear stress
appears to be similar to a multi-phase system with the fluid phase
``immersed'' in the solid phase. The fluid phase behaves as
a simple Newtonian fluid for small shear rates when the density is
almost constant, but exhibits shear thinning at larger shear rates
when the density begins to drop. This regime can be described by the
generalization of the theory \cite{note1} which includes the equations
for density and granular temperature following from the kinetic theory
of dilute granular gases. Our simulations were limited to 2D geometry.
While we anticipate that the structure of the model should
remain unchanged in 3D systems, the specific form of the fitting
functions may vary.
The authors are indebted to J. Gollub, P. Cvitanovic, T. Halsey, B. Beringer
for useful discussions, and to J.C. Tsai for sharing his unpublished
experimental data.  This work was supported by the Office of the Basic
Energy Sciences at the US DOE, grants W-31-109-ENG-38,
and DE-FG03-95ER14516.  Simulation were performed at the National Energy Research
Scientific Computing Center.

\end{document}